\documentclass[aps,pra,twocolumn,superscriptaddress,showpacs,showkeys,amsmath,amssymb]{revtex4-1}
\usepackage{amsmath}
\usepackage{accents}
\usepackage{bm}
\usepackage[english]{babel}
\usepackage{graphicx}
\usepackage[usenames]{color}
\usepackage{colortbl}
\usepackage{mathtools}
\usepackage{commath}
\usepackage[font=footnotesize,figurewithin=none]{caption}
\usepackage[caption=false]{subfig}
\usepackage[toc,page,title]{appendix}

\begin{document}

\title{Physical Realization of the {\L}uczka Model: Inducing Interaction and Entanglement Between Spins}

\author{A. R. Kuzmak}
\email{andrijkuzmak@gmail.com}
\affiliation{Department for Theoretical Physics, Ivan Franko National University of Lviv,\\ 12 Drahomanov St., Lviv, UA-79005, Ukraine}
\affiliation{Haiqu, Inc., 95 Third Street, San Francisco, CA 94103, USA}

\date{\today}
\begin{abstract}
The {\L}uczka model describes the interaction of spins with a common bosonic environment in the pure-dephasing regime, in which the bosonic modes couple directly to only one spin component. Due to its specific form of interaction,
the model is exactly solvable and has been widely used to describe spin decoherence and effective spin-spin interactions mediated by a common bosonic environment. However, to the best of our knowledge, its physical realization
and the specific conditions under which such an interaction can be implemented have not yet been addressed. Here, we propose a physical realization of the {\L}uczka spin-environment interaction by coupling the spin system
to unidirectionally propagating, polarized electromagnetic radiation. In contrast to the original {\L}uczka model, which allows an arbitrary angular distribution of photon wave vectors, our scheme gives
a unidirectional photon flux with retaining a frequency distribution of the photons. As an illustrative application of the proposed realization, we investigate spin-entanglement dynamics for coherent, single-mode thermal,
and Ohmic bosonic environments, examining their dependence on the number of spins in the system.
\end{abstract}

\maketitle

\section{Introduction \label{sec1}}

One of the exactly solvable models describing the interaction of a spin with a bosonic environment is the model proposed by {\L}uczka in 1990 \cite{luczka990}. This model describes the dynamics of a spin-1/2 system interacting
with noninteracting bosonic oscillators, whose state is described by the Gibbs distribution at temperature $T$. Because the interaction between the spin and each oscillator is restricted to a specific direction,
the resulting dynamics of the system can be solved exactly and described analytically. This makes it possible to investigate the dynamics of the spin under the influence of the environment, including its decoherence
and other physical properties \cite{Daika2008,Daika2010,Daika20102,Morozov2012,Daika2012,Daika20122,Ignatyuk2013,Ignatyuk2015,Ignatyuk2022,Ignatyuk2025}. It is worth noting that a detailed study of the decoherence dynamics
of this model, taking into account initial qubit-environment correlations, was first presented in \cite{Morozov2012} and was later extended to different types of environmental spectral density distributions
in subsequent papers \cite{Ignatyuk2013,Ignatyuk2015,Ignatyuk2022,Ignatyuk2025}.

One important consequence of the interaction between spins and a common environment is the emergence of an effective spin-spin interaction, which can ultimately lead to entanglement between the spins.
Oh and Kim were among the first to demonstrate that a common environment can mediate an effective Ising interaction between two spins  \cite{Oh2006}. They further investigated the resulting entanglement
dynamics for an environment characterized by an Ohmic spectral density. The interplay between entanglement and decoherence in qubit and qudit systems coupled to a common environment has subsequently been investigated
in many papers \cite{Privman2006,Privman2007,Daika20082,ContrerasPulido2008,Chaudhry2013,Tan2015}. More recently, the same mechanism and the resulting entanglement dynamics under different environmental conditions have been studied
using exactly solvable pure-dephasing models \cite{Mutahar2022,Mirza2024,Kuzmak2025,Smetana2026}. Particularly relevant to the present work, Mirza and Chaudhry \cite{Mirza2024} explicitly identified the interaction-generating term
in the evolution operator, demonstrating the emergence of an effective Ising interaction mediated by the common environment.

Although the {\L}uczka model is exactly solvable, and the dephasing and environment-mediated interactions it describes in spin systems have been extensively studied in recent years, to the best of our knowledge,
no physical realization of the model or explicit conditions under which the required spin-environment interaction can be realized have been proposed. In previous studies, the model has generally been treated
as an exactly solvable theoretical model owing to the specific form of the spin-environment coupling, without addressing its physical implementation. In the present work, we propose a physical scheme that realizes
this interaction using a unidirectionally polarized photon flux, thereby providing a possible physical implementation of the {\L}uczka model. In contrast to the original {\L}uczka model, which allows an arbitrary
angular distribution of photon wave vectors, our scheme assumes a unidirectional photon flux, while retaining the frequency distribution of the radiation. Such a configuration can be realized using laser radiation.
As an illustrative example, for an arbitrary number of spins, we investigate the dynamics of entanglement induced by environments described by a pure coherent state, corresponding to laser radiation, a single-mode thermal state,
and an environment characterized by an Ohmic spectral density.

The paper is organized as follows. In Sec.~\ref{scheme}, we derive the spin-photon interaction Hamiltonian and show how a suitable polarization of the photon flux leads to the {\L}uczka-type spin-environment interaction.
In Sec.~\ref{sevolution}, we derive the exact evolution operator and the reduced density matrix of the spin subsystem for an arbitrary number of spins. In Sec.~\ref{spinsevolutionent}, we investigate the resulting
entanglement dynamics for different photonic environments, including coherent, one-mode thermal, and Ohmic thermal environments. Finally, Sec.~\ref{concl} summarizes the main results and conclusions of the work.

\section{Spin-Photon Interaction Hamiltonian \label{scheme}}

We consider a system of noninteracting spins placed in an electromagnetic field. The spins interact only with the magnetic component of the field, which is confined to a periodic cubic volume $V$ and described by a vector potential
\begin{eqnarray}
{\bf A}({\bf r})=\sum_{{\bf k}}\sum_{\mu=1,2}\sqrt{\frac{2\pi c^2}{\omega_k V}}{\bf e}_{{\bf k},\mu}\left(b_{{\bf k,\mu}}e^{i({\bf k},{\bf r})}+b^\dagger_{{\bf k,\mu}}e^{-i({\bf k},{\bf r})}\right),\nonumber\\
\label{elmagnpot}
\end{eqnarray}
where $b^\dagger_{{\bf k},\mu}$ and $b_{{\bf k},\mu}$ are the creation and annihilation operators of a photon with wave vector ${\bf k}$ and polarization $\mu$, the radius vector ${\bf r}$ determines the spatial position of the field,
$\omega_k$ is the photon frequency corresponding to wave number $k$, ${\bf e}_{{\bf k},\mu}$ is the photon polarization vector, and $c$ is the speed of light. We set $\hbar=1$, which implies that energy is measured in units of frequency.
Taking into account that the magnetic field is given by the curl of the vector potential, ${\bf B}=\nabla \times \mathbf{A}(\mathbf{r})$, the Hamiltonian describing the interaction between the magnetic field and spatially separated spins located at positions
${\bf r}_i$ and the magnetic field takes the form
\begin{equation}
H_{int} = \gamma \sum_{i=1}^{N} \left(\mathbf{S}_i, \left[\nabla, \mathbf{A}(\mathbf{r}_i)\right] \right),
\label{hamiltonian}
\end{equation}
\begin{figure}[!!h]
\centerline{\includegraphics[scale=0.4, angle=0.0, clip]{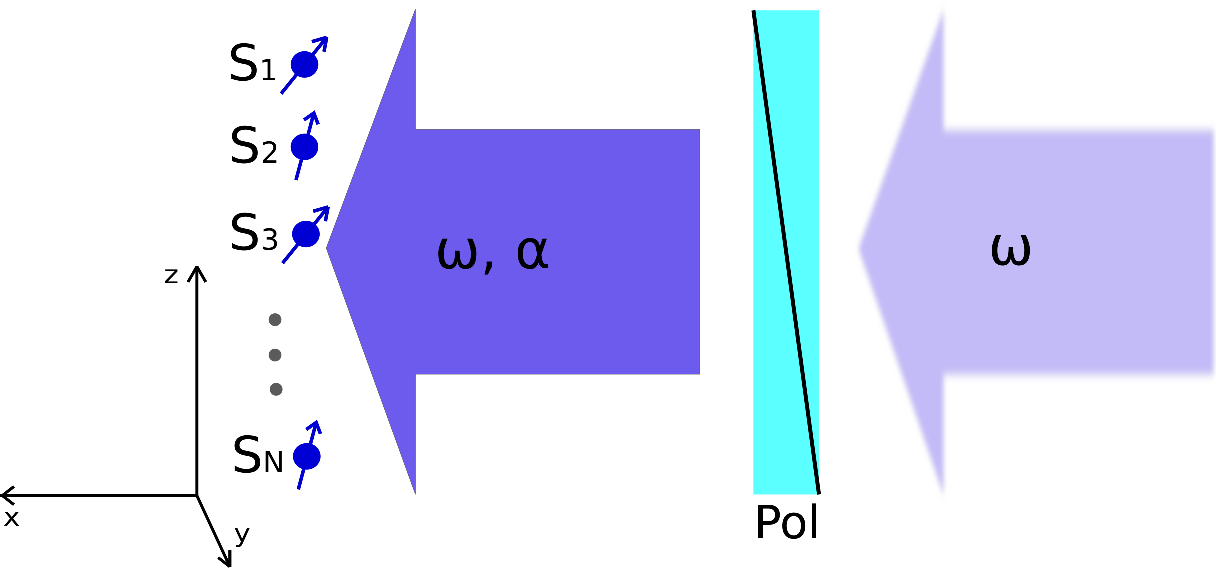}}
\caption{Schematic illustration of the physical realization of the spin-photon interaction Hamiltonian. Photons propagate along the $x$-axis, while the spins are aligned along the $z$-axis and distributed in a plane perpendicular
to the propagation direction. The radiation field is assumed to uniformly cover the entire spin ensemble. A polarizer placed in the propagation path selects a single polarization mode, resulting in the coupling of one spin component
to the photon field and thereby realizing the {\L}uczka-type interaction model.}
\label{model}
\end{figure}
where $\gamma$ is the gyromagnetic ratio of the spin.
In general, radiation consists of a collection of photons described by a spectral distribution that depends not only on the magnitude of the wave vector but also on the spatial distribution.
Each photon is therefore characterized by its wave vector and the corresponding polarization vectors. 
In the spherical coordinate system, these vectors can be written as
\begin{equation}
\begin{aligned}
&{\bf k}=k(\sin\theta_k\cos\phi_k, \sin\theta_k\sin\phi_k, \cos\theta_k),\\
&{\bf e}_{{\bf k}, 1}=(\cos\theta_k\cos\phi_k, \cos\theta_k\sin\phi_k, -\sin\theta_k),\\
&{\bf e}_{{\bf k},2}=(-\sin\phi_k,\cos\phi_k,0),
\label{vectors}
\end{aligned}
\end{equation}
where $\theta_k$ and $\phi_k$ are the polar and azimuthal angles specifying the direction of photon propagation.
Then using the fact that $\nabla\exp{\left(\pm i({\bf k},{\bf r_i})\right)}=\pm i {\bf k}\exp{\left(\pm i({\bf k},{\bf r_i})\right)}$ and $[{\bf k},{\bf e}_{{\bf k}, 1}]=k{\bf e}_{{\bf k}, 2}$,
$[{\bf k},{\bf e}_{{\bf k}, 2}]=-k{\bf e}_{{\bf k}, 1}$, Hamiltonian \eqref{hamiltonian} takes the form
\begin{widetext}
\begin{equation}
H_{int} = i \gamma \sum_{{\bf k}} \sqrt{\frac{2\pi\omega_k}{V}} \sum_{i=1}^{N} \left((\mathbf{S}_i,{\bf e}_{{\bf k},2})\left(b_{{\bf k},1}e^{i({\bf k},{\bf r}_i)}-b^\dagger_{{\bf k},1}e^{-i({\bf k},{\bf r}_i)}\right) -
(\mathbf{S}_i,{\bf e}_{{\bf k},1})\left(b_{{\bf k},2}e^{i({\bf k},{\bf r}_i)}-b^\dagger_{{\bf k},2}e^{-i({\bf k},{\bf r}_i)}\right)  \right).
\label{hamiltonian2}
\end{equation}
\end{widetext}
Let us assume that all photons propagate in the same direction. We choose the coordinate system such that this direction corresponds to $\theta_k = \pi/2$ and $\phi_k = 0$, i.e., along the $x$-axis.
In this case, the wave vector and the polarization vectors are identical for all photons and take the form $\mathbf{k} = k(1,0,0)$, $\mathbf{e}_{\mathbf{k},1} = (0,0,-1)$, $\mathbf{e}_{\mathbf{k},2} = (0,1,0)$.
We further assume that all spins are localized along the $z$-axis, which corresponds to a line perpendicular to the propagation direction of the radiation. The radiation field is assumed to be sufficiently wide
to uniformly cover the entire spatial region occupied by the spins. Since the photons propagate along the $x$-axis, the spins are therefore distributed in the $y$-$z$ plane, which implies $x_i = 0$.
Consequently, $(\mathbf{k},\mathbf{r}_i)=0$, which reduces the phase factors to unity, i.e., $e^{\pm i (\mathbf{k},\mathbf{r}_i)}=1$. The interaction Hamiltonian \eqref{hamiltonian2} then simplifies to
\begin{equation}
\begin{aligned}
&H_{int} = i\gamma \sum_{k} \sqrt{\frac{2\pi\omega_k}{V}}\\
&\sum_{i=1}^{N} \left(S_i^{y}\left(b_{k,1}-b^\dagger_{k,1}\right)+S_i^{z}\left(b_{k,2}-b^\dagger_{k,2}\right)\right).
\label{hamiltonian3}
\end{aligned}
\end{equation}
If a polarizer is placed in the propagation path of the radiation such that only one polarization mode is transmitted, the interaction reduces to a coupling between a single spin component and the photon field,
corresponding to the {\L}uczka-type model \cite{luczka990}. For definiteness, let us assume that the polarizer transmits only photons with polarization type 2. The interaction Hamiltonian then takes the form
\begin{equation}
H_{int} = \gamma \sum_{i=1}^{N} S_i^{z}\sum_{k}\left(g^*_{k}b_{k}+g_{k}b^\dagger_{k}\right),
\label{hamiltonian4}
\end{equation}
where $g_k=-i \sqrt{2\pi\omega_k/V}$. Since only a single polarization mode contributes to the interaction, the polarization index is omitted. The complete experimental implementation scheme for this interaction is shown in Fig.~\ref{model}.

\section{System dynamics \label{sevolution}}

Let us now consider the evolution of the complete system consisting of spins and photons, whose interaction is described by the Hamiltonian~(\ref{hamiltonian4}). The total Hamiltonian takes the form
\begin{equation}
H=H_{ph}+H_{int},
\label{totalhamiltonian}
\end{equation}
where $H_{ph}=\sum_k\omega_k b_k^{\dagger}b_k$ is the Hamiltonian of photonic subsystem. The evolution operator of the system can be written as (see Appendix~\ref{densitymatrix1})
\begin{widetext}
\begin{align}
&\exp{\left(-i(H_{ph}+H_{int})t\right)}\nonumber\\
&=\prod_k\exp{\left[\frac{\gamma}{\omega_k}\left( (e^{-i\omega_kt}-1)g_k b_k^\dagger - (e^{i\omega_kt}-1)g_k^* b_k\right)\sum_{i} S_i^z\right]}\exp{\left[-i\frac{\vert g_k\vert^2\gamma^2}{\omega_k^2}\left(\sin(\omega_kt)-\omega_kt\right)\left(\sum_{i} S_i^z\right)^2\right]}\nonumber\\
&\exp{\left[-i\omega_kb_k^\dagger b_kt\right]}.
\label{evolutionoperator}
\end{align}
\end{widetext}
The first exponential term describes the evolution arising from the coupling between the spins and the environment. It corresponds to a spin-dependent displacement of the environmental modes, whereby each spin configuration
induces a different coherent state of the bath, and can be viewed as an interaction with an effective magnetic field directed along the $z$-axis. The resulting factor gives rise to dephasing of the spin subsystem,
reflecting the loss of phase coherence caused by the effective magnetic field. As a result, different spin states become entangled with different environmental states, leading to decoherence after tracing out the bath degrees of freedom.
The second term describes an effective Ising evolution induced by the environment. The last term describes the free evolution of the environmental bosonic modes.

It is worth noting that the effective interaction between spins defined by parameter $2\sum_k\frac{\vert g_k\vert^2\gamma^2}{\omega_k^2}\left(\sin(\omega_kt)-\omega_kt\right)/t$, which rapidly reaches
a negative value $-2\sum_k\frac{\vert g_k\vert^2\gamma^2}{\omega_k}$. The induced interaction takes the form of ferromagnetic Ising (Curie-Weiss) model. For such a model, magnetic ordering of the spins can be observed.
Such magnetic ordering has been studied for spin systems coupled to a common bosonic environment in paper \cite{Min2024}. The authors demonstrated that a global bosonic bath can drive a quantum phase transition to the ordered phase.
As a results, bath-induced magnetic ordering can be measured.

Let us assume that, at the initial moment of time, all spins are directed in the same direction, so that each spin is described by the pure state $a\vert 0\rangle + b\vert 1\rangle$, while the photonic subsystem is described
by the density matrix $\rho_{ph}(0)$, where $a$ and $b$ are complex parameters satisfying the normalization condition $\vert a\vert^2+\vert b\vert^2=1$. It is worth noting that a such initial state of spins can be prepared
by applying a strong external magnetic field. Under these conditions, this state corresponds to either the ground or the excited state of the spin system.
The evolution of the whole system is derived in Appendix~\ref{densitymatrix2}. By tracing over the photonic degrees of freedom, we obtain the reduced evolution of the spin subsystem resulting from its interaction with the photon field.
Here, we present the reduced density matrix describing the spin subsystem, which can be expressed as

\begin{widetext}
\begin{align}
&\rho_s(t)=\sum_{n,n'=0}^N[\rho_s(0)]_{n,n'}\exp{\left[-i\frac{\Delta(t)}{4}\left((N-2n)^2-(N-2n')^2\right)\right]}\nonumber\\
&\left\langle \prod_k \exp\left[-(n-n')\frac{\gamma}{\omega_k}\left((e^{-i\omega_kt}-1)g_kb_k^{\dagger}-(e^{i\omega_kt}-1)g_k^*b_k\right)\right]\right\rangle_{\rho_{ph}(t)},
\label{Trphotons3}
\end{align}
\end{widetext}
where
\begin{align}
&[\rho_s(0)]_{n,n'}=a^{N-n}{a^*}^{N-n'}b^{n}{b^*}^{n'}\nonumber\\
&\times\sum_{j,j'}P_j\left(\vert 0\rangle^{\otimes(N-n)}\vert 1\rangle^{\otimes n}\right) P_{j'}\left(\langle 0\vert^{\otimes(N-n')}\langle 1\vert^{\otimes n'}\right)
\label{denotinsdm}
\end{align}
is the part of the initial spin density matrix corresponding to the fixed values of $n$ and $n'$, and
\begin{align}
\Delta(t)=\sum_k\frac{\vert g_k\vert^2\gamma^2}{\omega_k^2}\left(\sin(\omega_kt)-\omega_kt\right)
\label{deltaintpar}
\end{align}
is the time-dependent parameter that determines the strength of the interaction between spins multiplied by time $t$. Here $P_j$ runs over all distinct permutation operators generating the $\binom{N}{n}$ different arrangements of $n$ excitations
among $N$ qubits. The state $\sum_{j}P_j\left(\vert 0\rangle^{\otimes(N-n)}\vert 1\rangle^{\otimes n}\right)$ is an eigenstate of the $z$-component of the total spin operator $\sum_i S_i^z$ with the eigenvalue $(N-2n)/2$.
From Eq.~\eqref{Trphotons3}, one can identify two distinct contributions to each matrix element of the density matrix. The first part describes the unitary evolution of the spin subsystem, whereas the second describes the decoherence
of the spin subsystem. Note that the the averaging is performed with respect to the time-dependent density matrix of the photonic subsystem
$\rho_{ph}(t)=e^{-iH_{ph}t}\rho_{ph}(0)e^{iH_{ph}t}$.  In the next section, we investigate this evolution for different initial states of the photonic environment.

\section{Entanglement dynamics of the spins in different photonic environments \label{spinsevolutionent}}

To avoid presenting large formulas here, we give the density matrix for the case in which the initial state of the spins is projected along the $x$-axis, i.e., each spin is in the state
$\vert +\rangle =1/\sqrt{2}(\vert 0\rangle + \vert 1\rangle)$ ($a=b=1/\sqrt{2}$). This choice is motivated by the fact that, for a projection of the spins perpendicular to the direction of the induced interaction,
maximum entanglement between them can be achieved (see, for example, Appendix~C in \cite{Smetana2026}).

We now consider the evolution of the spin subsystem for different initial states of the photonic subsystem. We investigate the entanglement in such a system. As an example, we consider the entanglement between one spin and the remaining
spins in the system as a function of the total number of spins $N$. For this purpose, we use negativity as a measure of entanglement \cite{Vidal2002,Plenio2005}.
Thus, the entanglement between subsystems $A$ and $B$ can be characterized by the sum of the negative eigenvalues $\Lambda_i$ of the partially transposed density matrix $\rho^{{\rm \Gamma}_{A(B)}}$ with respect to subsystem $A(B)$.
It can be expressed as follows

\begin{eqnarray}
\mathcal{N}(\rho)=\left\vert \sum_{\Lambda_i<0}\Lambda_i\right\vert=\sum_{i}\frac{\vert\Lambda_i\vert-\Lambda_i}{2}.
\label{negativity}
\end{eqnarray}

\subsection{Coherent-state environment \label{coherentenv}}

The quantum state of an ideal laser field with single mode is well approximated by a coherent state \cite{Glauber1963}, which is defined as the eigenstate of the annihilation operator
\begin{equation}
b \vert\alpha\rangle_{ph} = \alpha \vert\alpha\rangle_{ph},
\end{equation}
where $\alpha$ is the eigenvalue of operator $b$, which in general can be a complex number. Here, we omit the subscripts on the bosonic operators, since only a single bosonic mode is considered.
In the Fock-state basis, the coherent state is written as
\begin{equation}
\vert\alpha\rangle_{ph} =e^{-\vert\alpha\vert^2/2} \sum_{N_k=0}^{\infty} \frac{\alpha^{N_k}}{\sqrt{N_k!}}\vert N_k\rangle.
\label{coherentstate}
\end{equation}
For this initial state of photons the average in Eq.~\eqref{Trphotons3} is easy calculated and takes the form
\begin{align}
&\langle\ldots\rangle_{\rho_{ph}(t)}\nonumber\\
&=\exp{\left[-(n-n')^2\frac{4\gamma^2\vert g\vert^2}{\omega^2}\sin^2\left(\frac{\omega t}{2}\right)\left(\alpha^2e^{\alpha^2-\vert\alpha\vert^2}+1/2\right)\right]}\nonumber\\
&=\exp{\left[-4(n-n')^2\Gamma(t)\right]},
\label{Trphoroncoherent}
\end{align}
where we introduce the notation of the decoherence parameter $\Gamma(t)=\frac{\gamma^2\vert g\vert^2}{\omega^2}\sin^2\left(\frac{\omega t}{2}\right)\left(\alpha^2e^{\alpha^2-\vert\alpha\vert^2}+1/2\right)$.
Here and in what follows, we use the identity $\left\langle \exp{\left[z_1 b_k^\dagger+z_2 b_k\right]} \right\rangle_{\rho_{ph}(t)}=\exp[z_2z_1(\langle b_k^\dagger b_k\rangle_{\rho_{ph}(t)}+1/2)]$, where, for state
given by Eq.~\eqref{coherentstate}, $\langle b_k^+b_k\rangle_{\rho_{ph}(t)}=\alpha^2e^{\alpha^2-\vert\alpha\vert^2}$.

\begin{figure*}[t]
\centerline{\includegraphics[scale=0.34, angle=0.0, clip]{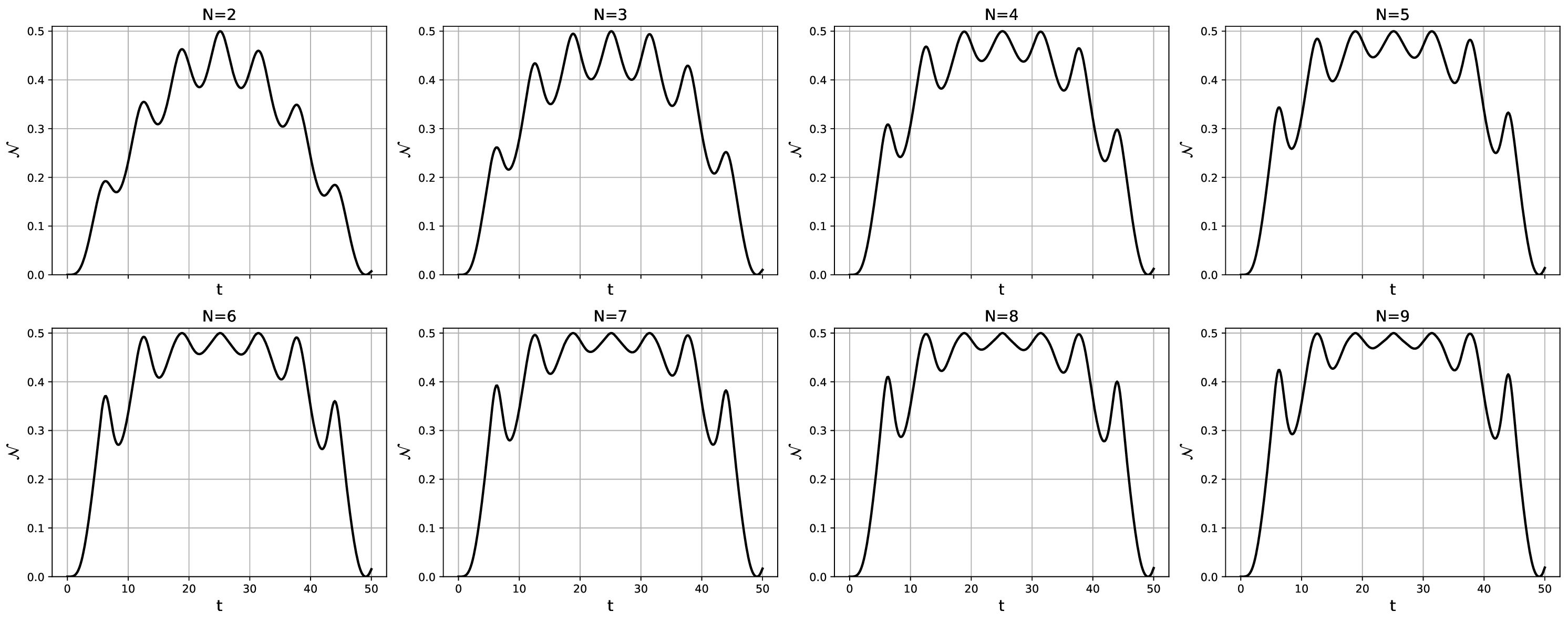}}
\caption{Time dependence of the entanglement between a single spin and the remaining spins for an environment described by the coherent state \eqref{coherentstate},
with system parameters $\gamma=1$, $\omega=1$, $\alpha=0$, and $V=100$, starting from states in which all spins are directed along the $x$-axis. The dependencies are shown for different numbers of spins.}
\label{negativity_coherence}
\end{figure*}

\begin{figure}[h]
\centerline{\includegraphics[scale=0.38, angle=0.0, clip]{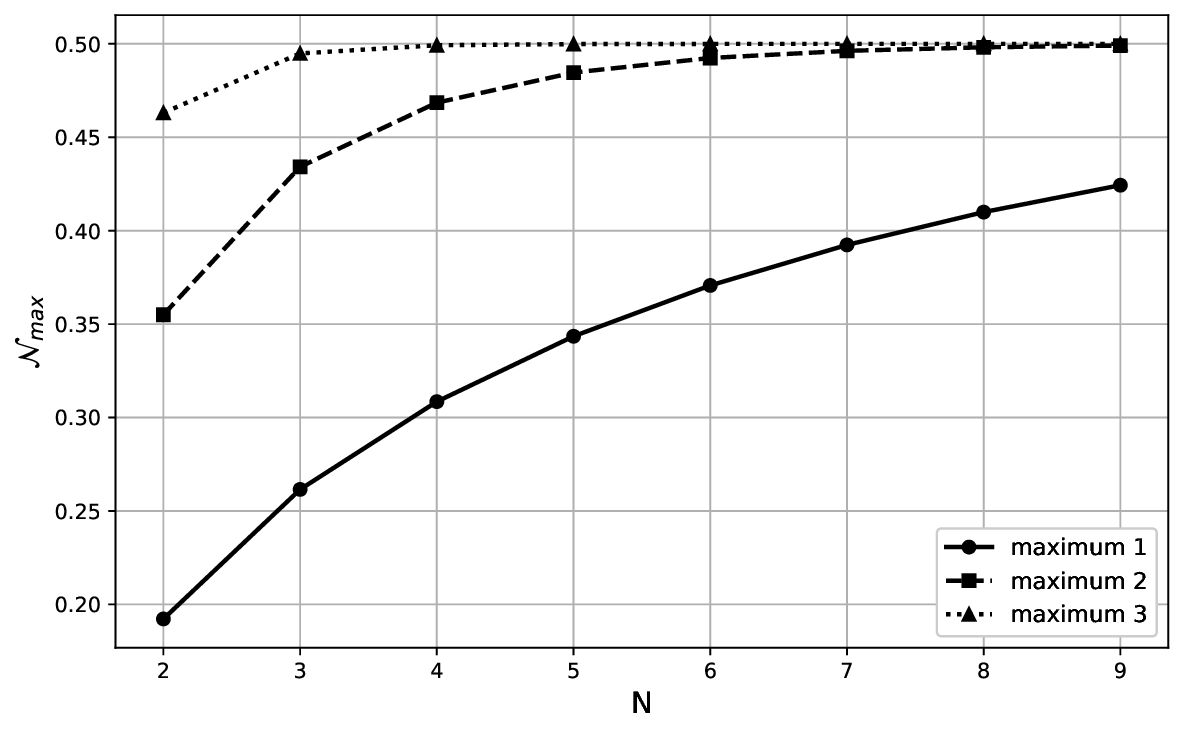}}
\caption{Maximum entanglement values at the first, second, and third peaks of the time evolution shown in Fig.~\ref{negativity_coherence}, as a function of the number of spins in the system.}
\label{negativity_coherence2}
\end{figure}

Substituting the average value \eqref{Trphoroncoherent} into the spin density matrix \eqref{Trphotons3} and assuming that initially all spins are separable and defined by the parameters $a=b=1/\sqrt{2}$,
we numerically calculate the time evolution of the negativity as a function of the number of spins $N$. As an example, Fig.~\ref{negativity_coherence} shows the time evolution of the negativity between one spin
and the remaining spins for $N=2$ to $9$, for a single-mode coherent photonic environment reduced to the vacuum limit $\alpha=0$, with $\gamma=1$, $\omega=1$, and $V=100$. Since $\alpha=0$ corresponds
to the photonic bath being initially in its vacuum state, the induced decoherence defined by the parameter $\Gamma(t)$ in Eq.~\eqref{Trphoroncoherent} is reduced to its minimum, while the induced Ising phase $\Delta(t)$
does not depend on $\alpha$.

The peaks in Fig.~\ref{negativity_coherence} arise from the structure of $\Gamma(t)$ in Eq.~\eqref{Trphoroncoherent}. Since $\Gamma(t) \propto \sin^2(\omega t/2)$, the decoherence parameter vanishes exactly at moment of time
$t_m = 2\pi m/\omega$ ($m = 0, 1, 2, \dots$). At these moments, the spin subsystem is protected from decoherence, and the entanglement is governed solely by the Ising phase $\Delta(t)$.

Figure~\ref{negativity_coherence2} shows the dependence of the first three local maxima of the negativity ($\mathcal{N}_{max}$) on the number of spins $N$, obtained from the curves in Fig.~\ref{negativity_coherence}.
The negativity increases with the number of spins. This can be attributed to the fact that the induced Ising interaction couples all spins to each other, so that increasing the number of spins leads to stronger entanglement
between an individual spin and the rest of the system.

\subsection{Thermal environment \label{thermalenv}}

\begin{figure*}[t]
\centerline{\includegraphics[scale=0.34, angle=0.0, clip]{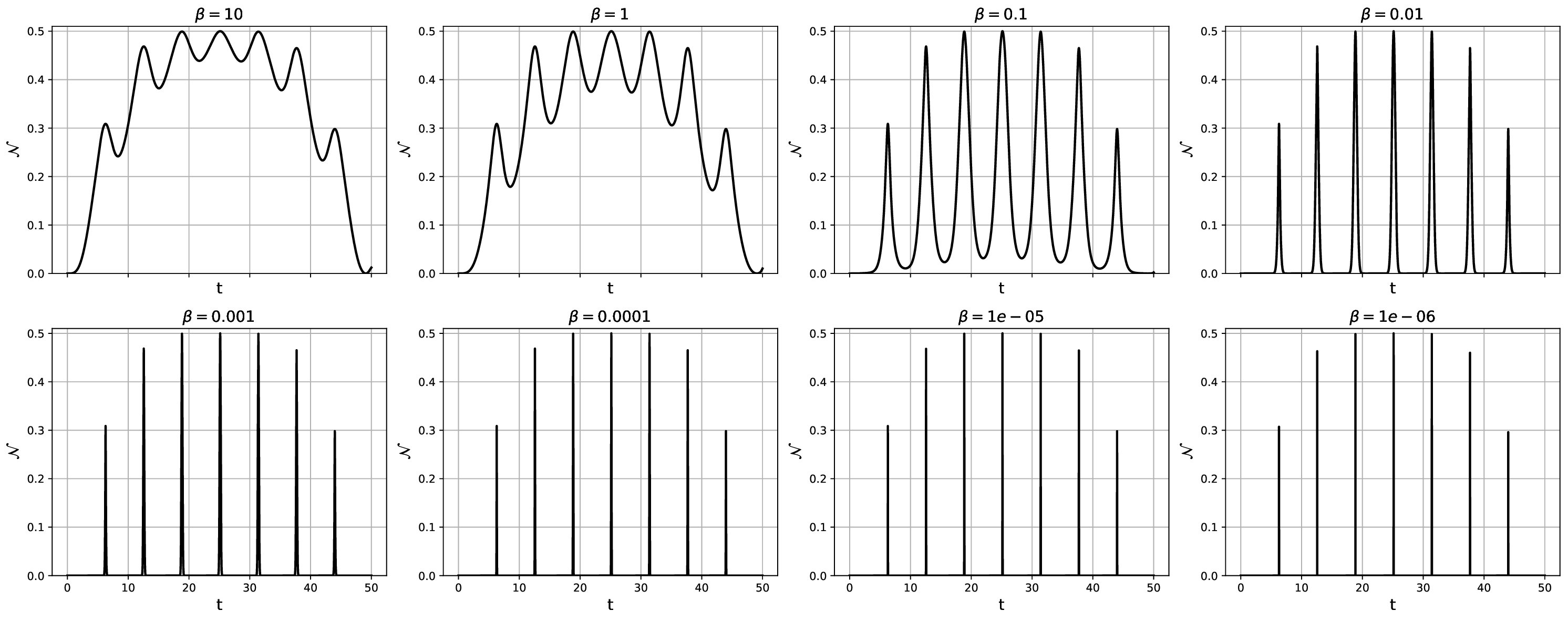}}
\caption{Time dependence of the entanglement between a single spin and the remaining spins for a single-mode thermal photonic environment described by the state \eqref{thermalstatedm}, with $N=4$, $\gamma=1$, $\omega=1$, and $V=100$, starting from a state in which all spins are directed along the $x$-axis, for different temperatures.}
\label{negativity_onemode_thermal}
\end{figure*}

\begin{figure*}[t]
\centerline{\includegraphics[scale=0.34, angle=0.0, clip]{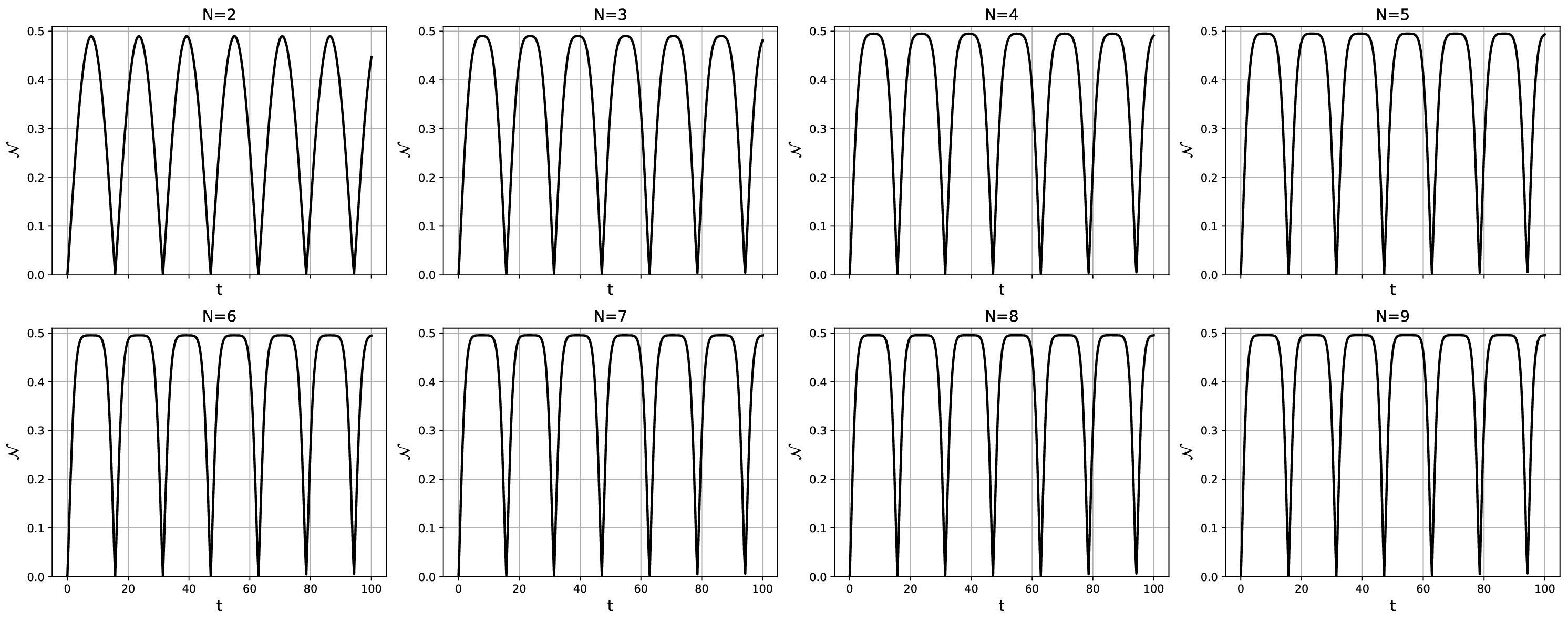}}
\caption{Time dependence of the entanglement between a single spin and the remaining spins for a thermal photonic environment characterized by the Ohmic spectral density \eqref{spectral}, with system parameters $\gamma=1$, $\omega_c=10$, $\lambda=0.01$, $s=3$, and $\beta=1$, starting from states in which all spins are directed along the $x$-axis.
The dependencies are shown for different numbers of spins.}
\label{negativity_ohmic}
\end{figure*}

In the case of a thermal photonic environment, its quantum state in thermodynamic equilibrium is described by the Boltzmann distribution
\begin{align}
\rho_{ph}=\frac{1}{Z_{ph}}e^{-\beta H_{ph}},
\label{thermalstatedm}
\end{align}
where
\begin{align}
Z_{ph}={\rm Tr}e^{-\beta \sum_k\omega_kb_k^{\dagger}b_k}=\prod_k\frac{1}{\left(1-e^{-\beta\omega_k}\right)}
\end{align}
is a partition function of the photonic environment, and $\beta=1/T$ is the inverse temperature in the units when the Boltzmann constant $k_B=1$. Then using the fact that
$\langle b_k^{\dagger}b_k\rangle_{\rho_{ph}}=1/(e^{\beta\omega_k}-1)$, the mean value in the equation \eqref{Trphotons3} takes the form
\begin{align}
\langle\ldots\rangle_{\rho_{ph}(t)}=\exp{\left[-2\left(n-n'\right)^2\Gamma(t)\right]},
\label{Trphoronthermal2}
\end{align}
where the decoherence parameter $\Gamma(t)=\sum_k\frac{\vert g_k\vert^2\gamma^2}{\omega_k^2}\sin^2\left(\frac{\omega_kt}{2}\right)\coth\left(\beta\omega_k/2\right)$. It is easy to see that at very low temperatures
($\coth\left(\beta\omega_k/2\right)=1$) we obtain the result for the case of the vacuum state of a coherent-state environment (see, equation \eqref{Trphoroncoherent} for $\alpha=0$).

In this case, the entanglement is additionally reduced by the temperature of the environment. Let us first consider a single-mode environment at different temperatures.
As shown in Fig.~\ref{negativity_onemode_thermal}, for $N=4$, $\gamma=1$, $\omega=1$, and $V=100$, the negativity is calculated for a sequence of inverse temperatures $\beta = 10,\ 1,\ 10^{-1},\ 10^{-2},\ 10^{-3},\ 10^{-4},\ 10^{-5},\ 10^{-6}$.
The height of the negativity maxima at the moment of times $t_m = 2\pi m/\omega$ ($m=1,2,\dots$) remains the same for all values of $\beta$. At these times, $\sin^2(\omega t_m/2) = 0$, so that $\Gamma(t_m) = 0$ regardless
of temperature. Therefore, the negativity at these points is determined only by the temperature-independent phase $\Delta(t)$.
Away from these moments of time, the decoherence parameter $\Gamma(t)$ increases with temperature through the factor $\coth\left(\beta\omega/2\right)$, which diverges as $\beta \to 0$. As a result, increasing the temperature suppresses
the negativity more strongly away from the revival times. Thus, the broad maxima observed at low temperature ($\beta = 10,,1$) become progressively narrower and turn into sharp peaks at the moments of time $t_m = 2\pi m/\omega$
for high temperature ($\beta \lesssim 10^{-3}$), as shown in Fig.~\ref{negativity_onemode_thermal}. Therefore, increasing the temperature does not change the maximum values of the entanglement, but makes the entanglement significant
only in a narrow time interval around the decoherence-free times $t_m = 2\pi m/\omega$.

In the continuum limit, the discrete sum over field modes is replaced by an integral over the continuous frequency $\omega$, weighted by a spectral
density $J(\omega)$ that combines the mode density with the squared coupling strength $|g(\omega)|^2$ at each frequency, as is standard in spin-boson-type models. Under this substitution,
\begin{align}
\sum_{k} \vert g_k\vert^2 f(\omega_k) = \int_0^{\infty} J(\omega) f(\omega) d\omega.
\label{spectraltransf}
\end{align}
As an example we consider the well-known Ohmic spectral density used in various spin-boson models \cite{Breuer2007,Morozov2012,Luczka2013,Ignatyuk2015,Ignatyuk2022,Barr2024,Ignatyuk2025}
\begin{equation}
J(\omega)=\lambda\omega_c^{1-s}\omega^s e^{-\omega/\omega_c},
\label{spectral}
\end{equation}
where $\lambda$ is the frequency-independent overall coupling strength of the spin-boson interaction, setting the amplitude of the spectral density $J(\omega)$, the index $s$ called the ohmicity parameter defined the nature of the environment ($0 < s < 1$ corresponds to the sub-Ohmic regime, $s = 1$ is determines the Ohmic
regime, and $s > 1$ corresponds to the super-Ohmic one), $\omega_c$ is the cut-off frequency, which means that $J(\omega)\to 0$ at $\omega\to\infty$.

Since the evolution of spin systems in an Ohmic environment has already been extensively studied, as an example here we consider a case in which the entanglement of the system reaches its
maximal possible value, and examine how it behaves with a change in the number of spins in the system. To this end, we consider a super-Ohmic environment defined by ohmicity parameter
$s=3$, cutoff frequency $\omega_c=10$, coupling strength $\lambda=0.01$, giromagnetic ratio of spin $\gamma=1$, and inverse temperature $\beta=1$. The corresponding time evolution of the negativity for $N=2$ to $9$ is shown in Fig.~\ref{negativity_ohmic}.
In contrast to the Ohmic case, where decoherence accumulates irreversibly and the negativity decays to zero (for exaple, see \cite{Tan2015,Smetana2026}), the super-Ohmic spectral density strongly suppresses
the bath modes responsible for decoherence vanishes rapidly. As a result, the induced Ising interaction parameter $\Delta(t)$ dominates over the decoherence parameter $\Gamma(t)$ for the evolution, and the negativity achieve
to its maximal value $\mathcal{N}=1/2$ as is shown in Fig.~\ref{negativity_ohmic}. As the number of spins increases, the induced interaction among all the spins widens the time window over which the maximal entanglement value
is sustained, so that the system remains in the maximally entangled state for longer period of time, as one would expect.

\section{Conclusions \label{concl}}

In this work, we have proposed a physical realization of the {\L}uczka model in the case where all wave vectors of the medium have the same direction. This is an exactly solvable model that describes the interaction of spins with
a common bosonic environment in the pure-dephasing regime. We have shown that this interaction can be implemented by coupling an ensemble of spins to a unidirectionally propagating, polarized photon flux, with the spins aligned
along the $z$-axis and localized in a plane perpendicular to the propagation direction of the radiation, while a polarizer selects a single polarization mode. The resulting spin-photon interaction Hamiltonian reduces exactly
to the {\L}uczka-type Hamiltonian. This realization corresponds to the special case of the {\L}uczka model in which all photon wave vectors are parallel, while their frequency distribution is retained. Such a unidirectional photon
flux can be realized by laser radiation, providing a direct physical route to the {\L}uczka-type spin-environment interaction.

One way of realizing such an interaction is by irradiating the spins with laser light passed through a polarizer. As a first example, we have therefore investigated the evolution of the entanglement between a single spin
and the remaining spins under the influence of an environment described by a coherent state. The time dependence of entanglement has been examined as a function of the number of spins in the system. Because the environment mediates
an effective interaction between all pairs of spins, it follows naturally that the degree of entanglement in the system increases as the number of spins grows. In addition, we have also studied the entanglement evolution in a single-mode thermal
environment, as well as in an environment characterized by an Ohmic spectral density.

\section{Acknowledgements}

This work was supported by Project 2025.07/0108 from National Research Foundation of Ukraine. We are grateful to Profs. Volodymyr Tkachuk for helpful advices.

\begin{appendices}

\section{Evolution operator \label{densitymatrix1}}
\setcounter{equation}{0}
\renewcommand{\theequation}{A\arabic{equation}}

In this appendix, we present the evolution operator of the entire system spin in a form convenient for further calculation \eqref{evolutionoperator}. Using Zassenhaus formula \cite{Magnus1954,Cassas2012}
\begin{align}
&\exp{\left(-i(H_{ph}+H_{int})t\right)}=\exp{(-iH_{ph}t)}\exp{(-iH_{sb}t)}\nonumber\\
&\times\exp{\left(\frac{t^2}{2!}[H_{ph},H_{int}]\right)}\nonumber\\
&\times\exp{\left(i\frac{t^3}{3!}([H_{ph},[H_{ph},H_{int}]]+2[H_{int},[H_{ph},H_{int}]])\right)}\ldots,
\label{Zassenhausformula}
\end{align}
and taking into account the commutation relations for bosonic operators $[b_k,b_{k'}^{\dagger}]=\delta_{kk'}$, $[b_k,b_k']=[b_k^{\dagger},b_{k'}^{\dagger}]=0$, the evolution operator can be expressed as a product
in which the free evolution of the photonic subsystem is explicitly separated from the interaction-induced dynamics
\begin{widetext}
\begin{align}
&\exp{\left(-i(H_{ph}+H_{int})t\right)}=\prod_k\exp{\left[-i\omega_kb_k^{\dagger}b_kt\right]}\exp{\left[\frac{\gamma}{\omega_k}(1-\cos(\omega_kt))\left(g_k b_k^{\dagger}-g_k^*b_k\right)\sum_iS_i^z\right]}\nonumber\\
&\times\exp{\left[-i\frac{\gamma}{\omega_k}\sin(\omega_kt)\left(g_k b_k^{\dagger}+g_k^*b_k\right)\sum_iS_i^z\right]}\exp{\left[i\frac{\vert g_k\vert^2\gamma^2}{\omega_k^2}\left(\omega_kt-\sin(\omega_kt)(2-\cos(\omega_kt))\right)\left(\sum_iS_i^z\right)^2\right]}.
\label{sepoperat}
\end{align}
\end{widetext}
Now applying the Baker-Campbell-Hausdorff formula, we replace the operator $\exp{\left[-i\omega_kb_k^+b_kt\right]}$ with other operators. Then the evolution operator takes the form
\begin{widetext}
\begin{align}
&\exp{\left(-i(H_{ph}+H_{int})t\right)}=\prod_k\exp{\left[\frac{\gamma}{\omega_k}(1-\cos(\omega_kt))\left(g_ke^{-i\omega_kt} b_k^{\dagger}-g_k^*e^{i\omega_kt}b_k\right)\sum_iS_i^z\right]}\nonumber\\
&\times\exp{\left[-i\frac{\gamma}{\omega_k}\sin(\omega_kt)\left(g_ke^{-i\omega_kt} b_k^{\dagger}+g_k^*e^{i\omega_kt}b_k\right)\sum_iS_i^z\right]}\exp{\left[i\frac{\vert g_k\vert^2\gamma^2}{\omega_k^2}\left(\omega_kt-\sin(\omega_kt)(2-\cos(\omega_kt))\right)\left(\sum_iS_i^z\right)^2\right]}\nonumber\\
&\times\exp{\left[-i\omega_kb_k^{\dagger}b_kt\right]}.
\label{operatorsfromappA}
\end{align}
\end{widetext}
The operator that is the result of commutator between operators from the first two exponents commutes with these operators. Therefore, applying the Zassenhaus formula \eqref{Zassenhausformula} with this fact
to equation \eqref{operatorsfromappA} yields the evolution operator in the form of equation \eqref{evolutionoperator}.

\section{Density matrix \label{densitymatrix2}}
\setcounter{equation}{0}
\renewcommand{\theequation}{B\arabic{equation}}

In this appendix, using the exact form of the evolution operator \eqref{evolutionoperator}, we obtain the density matrix of the entire system starting from a state where all spins are in the same state $a\vert 0\rangle+b\vert 1\rangle$,
and the density matrix of the photonic system has the form $\rho_{ph}(0)$. At the initial moment of time, the spins are separated from the photonic environment, and the density matrix has the form
\begin{widetext}
\begin{align}
&\rho(0)=\left(a\vert 0\rangle + b\vert 1\rangle\right)^{\otimes N}\left(a^*\langle 0\vert + b^*\langle 1\vert\right)^{\otimes N}\otimes\rho_{ph}(0)\nonumber\\
&=\sum_{n,n'=0}^{N}a^{N-n}{a^*}^{N-n'}b^{n}{b^*}^{n'}\sum_{j,j'}P_j\left(\vert 0\rangle^{\otimes(N-n)}\vert 1\rangle^{\otimes n}\right)P_j'\left(\langle 0\vert^{\otimes(N-n')}\langle 1\vert^{\otimes n'}\right)\otimes\rho_{ph}(0)
=\sum_{n,n'=0}^{N}[\rho_s(0)]_{n,n'}\otimes\rho_{ph}(0),
\label{initialwholedm}
\end{align}
\end{widetext}
where we introduce notation $[\rho_s(0)]_{n,n'}$ of the part of the spin density matrix defined by equation \eqref{denotinsdm}, $P_j$ runs over all distinct permutation operators generating the $\binom{N}{n}$
different arrangements of $n$ excitations among $N$ qubits. Using the fact that the state $\sum_{j}P_j\left(\vert 0\rangle^{\otimes(N-n)}\vert 1\rangle^{\otimes n}\right)$ is an eigenstate of the $z$-component of the total spin operator
$\sum_i S_i^z$ with the eigenvalue $(N-2n)/2$, we obtain the evolution of the density matrix of the whole system in the form
\begin{widetext}
\begin{align}
&\rho(t) = \sum_{n,n'=0}^N[\rho_s(0)]_{n,n'}\exp{\left[i\frac{\Delta(t)}{4}\left((N-2n)^2-(N-2n')^2\right)\right]}\nonumber\\
&\times\prod_k \exp\left[\frac{N-2n}{2}\frac{\gamma}{\omega_k}\left((e^{-i\omega_kt}-1)g_kb_k^{\dagger}-(e^{i\omega_kt}-1)g_k^*b_k\right)\right]\rho_{ph}(t)\nonumber\\
&\times\prod_k\exp\left[-\frac{N-2n'}{2}\frac{\gamma}{\omega_k}\left((e^{-i\omega_kt}-1)g_kb_k^{\dagger}-(e^{i\omega_kt}-1)g_k^*b_k\right)\right],
\label{wholedm}
\end{align}
\end{widetext}
where we introduce the notation $\rho_{ph}(t)=e^{-iH_{ph}t}\rho_{ph}(0)e^{iH_{ph}t}$ and the parameter $\Delta(t)$ is defined by equation \eqref{deltaintpar}.
To investigate the dynamics of the spin subsystem, we consider the reduced density matrix obtained by tracing out the photonic degrees of freedom This operation amounts to replacing, in Eq.~\eqref{wholedm},
the density matrix elements associated with the bosonic subsystem by their expectation values
\begin{widetext}
\begin{align}
&{\rm Tr}\left\{\prod_k\exp\left[(N-2n)\frac{\gamma}{2\omega_k}\left((e^{-i\omega_kt}-1)g_kb_k^{\dagger}-(e^{i\omega_kt}-1)g_k^*b_k\right)\right]\rho_{ph}(t)\right.\nonumber\\
&\left.\times\prod_k\exp\left[-(N-2n')\frac{\gamma}{2\omega_k}\left((e^{-i\omega_kt}-1)g_kb_k^{\dagger}-(e^{i\omega_kt}-1)g_k^*b_k\right)\right]\right\}\nonumber\\
&=\left\langle \prod_k \exp\left[-(n-n')\frac{\gamma}{\omega_k}\left((e^{-i\omega_kt}-1)g_kb_k^{\dagger}-(e^{i\omega_kt}-1)g_k^*b_k\right)\right]\right\rangle_{\rho_{ph}(t)},
\label{Trphotons2}
\end{align}
\end{widetext}
where the averaging is performed with respect to the density matrix $\rho_{ph}(t)$. By tracing over the photonic degrees of freedom, we obtain the reduced evolution of the spin subsystem \eqref{Trphotons3} arising
from its interaction with the photon field.

\end{appendices}

{}

\end{document}